\documentclass[twocolumn,showpacs,preprintnumbers,amsmath,amssymb]{revtex4}
\usepackage{graphicx}
\usepackage{dcolumn}
\usepackage{bm}
\usepackage[latin1]{inputenc} 
\newcommand{\comment}[1]{}

\begin{document}
\setlength{\unitlength}{0.7\textwidth} \preprint{}

\title{Rapid generation of angular momentum in bounded magnetized plasma}

\author{Wouter J.T.  Bos$^{1,2}$, Salah Neffaa$^{2}$ Kai Schneider$^{2}$}

\affiliation{$^1$ LMFA, UMR CNRS 5509, Ecole Centrale de Lyon -\\
Universit\'e de Lyon, Ecully, France}
\affiliation{$^2$ M2P2, UMR 6181 CNRS \& CMI, Universit\'es d'Aix-Marseille, Marseille, France}

\begin{abstract}

Direct numerical simulations of two-dimensional decaying MHD turbulence in 
bounded domains show the rapid generation of angular momentum in 
non-axisymmetric geometries. It is found that magnetic fluctuations 
enhance this mechanism. 
On a larger timescale, the generation of a magnetic angular momentum,
or angular field, is observed.
For axi-symmetric geometries, the generation of angular 
momentum is absent, nevertheless a weak angular field can be observed. The 
derived evolution equations for both angular momentum and angular field 
yield possible explanations for the observed behaviour.
\end{abstract}


\maketitle

The generation of large coherent structures of the size of the flow domain
is a generic feature of two-dimensional (2D) turbulence. Indeed, due to the
inverse energy cascade, 2D flows show a tendency to create space
filling structures. The nature of these structures and the way they are produced vary from flow to flow. In the context of Navier-Stokes turbulence the generation of a large-scale domain-filling structure was predicted by Kraichnan \cite{Kraichnan67} and  observed in the case of forced turbulence in a periodic domain in which energy condenses at the smallest possible wave number modes
 \cite{Lilly1969,Hossain1983}. 
In forced wall-bounded flows this was
reproduced numerically \cite{Heijst2006} and
experimentally \cite{Sommeria1986}, and it was shown that a large scale
rotating structure emerges, which dramatically reduces the level of the turbulent
fluctuations \cite{Shats2007}. 

A similar observation can be made in fusion plasmas, in which the
dynamics share many features with 2D flows due to the imposed magnetic
field. It is often assumed that
in these plasmas large scale poloidal structures, called zonal flows, are beneficial
for the confinement as they suppress turbulence and shear apart radially extended structures, which are largely responsible for anomalous transport \cite{Biglari1990,Terry2000,Diamond2005}. The hereby created
transport barriers might
play a key role in the transition to an improved
confinement state (H-mode)  \cite{Wagner1982}. In the case of MHD turbulence the role of rotation was shown to have a similar effect on the flow, reducing the velocity fluctuations and hereby stabilizing the magnetic field \cite{Shan1994}. In the present work we will continue the investigation of wall bounded non-ideal MHD. The generation of zonal flows through the absence of charge neutrality will not be addressed (charge neutrality being implied by the one-field MHD approximation). However, MHD allows for an affordable global description of non-uniform magnetoplasmas \cite{Montgomery1999}. 
The present work could be related to the L-H transition through the beneficial
effects of large scale poloidal rotation (which is observed in the present work) on the confinement of the
plasma. The present study is also motivated by the observation that
MHD-equilibria in toroidal geometry imply finite flow-fields due to
the finite viscosity and
resistivity \cite{Bates1998,Kamp2003,Montgomery1999}. In
these works non-ideal MHD steady states were investigated in both the
limit of small and large viscosity. In each case it was shown that the
steady state contains non-vanishing velocity fields, at odds with
classical static equilibria, on which decades of confinement research
are based. In the present work we will not consider steady states but
we will investigate the full nonlinear relaxation of non-ideal MHD
with non-trivial boundary conditions in two space dimensions. The resistivity and viscosity are non-zero but small, allowing for a turbulent flow. 
This approach can not take into account toroidal velocities and non-uniform toroidal magnetic fields and the extension of the present approach to three dimensions constitutes therefore an important direction for further research.

In the case of decaying Navier-Stokes turbulence it is shown that the self-organization in a
periodic domain will lead to a final state, consisting of two, non-interacting,
counter-rotating vortices  \cite{Joyce1973}. This picture changes however in the
presence of no-slip walls. In this case the flow relaxes to a state
with or without angular momentum, depending on the shape of the
domain \cite{Li1997,Clercx1998,Clercx2001}. Indeed in circular domains
without initial angular momentum the flow generally
relaxes to a state free from angular momentum \cite{Schneider2005-2}, whereas as soon as the
axi-symmetry is broken the flow relaxes to a state containing a domain filling
structure, containing significant angular momentum \cite{Keetels2008}. Theoretical progress has been made to explain the phenomenon in the inviscid case, based on a model of interacting vortices \cite{Pointin1976,Chavanis1996,Taylor2008}.

In the case of bounded two-dimensional MHD it is not known, up to now,
to which kind of state the flow relaxes and this will be addressed in
the present letter. We investigate the case in which both the magnetic
field and the velocity field can not penetrate into the walls. The
velocity field obeys the no-slip condition at the wall, whereas the
tangential component of the magnetic field can freely evolve, allowing
a net current through the domain. We will focus however in the present
study on the case in which no net current is initially present.

\begin{figure*}
\setlength{\unitlength}{1.\textwidth}
\includegraphics[width=1.\unitlength]{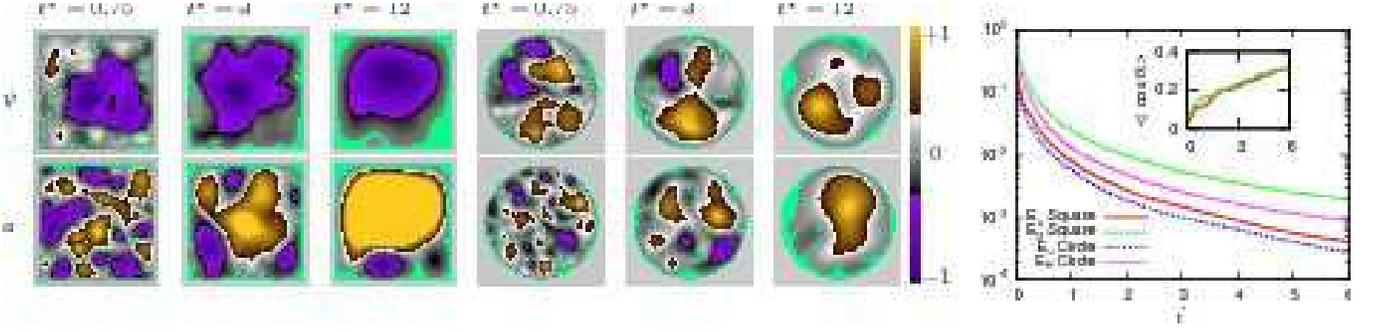}
\caption{(Color Online) Visualizations of the stream-function $\psi$ 
and the vector-potential $a$  
for both geometries. 
Figures for $\psi$ are normalized by the maximum of $|\psi|$ (and max($|a|$) for $a$). The values of max($|\psi|$) are from left to right $0.1$, $0.05$, $0.03$; $0.06$, $0.03$, $0.01$; and for max($|a|$) $0.08$, $0.06$, $0.04$;  $0.04$, $0.03$, $0.01$. Right: time-evolution of the kinetic and magnetic energy in both geometries. In the inset, the evolution of the absolute value of the relative cross-helicity $\left<|\cos(\theta)|\right>$, illustrates the global alignment of the velocity and magnetic field in both geometries. Red solid line: square geometry; green dashed line: circular geometry.
\label{visu}}
\end{figure*}

We start by writing the governing equations. In the present case we define
 two angular momenta: a kinetic and a magnetic one,
\begin{equation}\label{eqLu}
L_u=\int_{\Omega} \bm e_z\cdot (\bm r\times \bm u) dA,~~~L_B=\int_{\Omega} \bm e_z\cdot (\bm r\times \bm B) dA 
\end{equation}
in which ${\Omega}$ is the flow domain, $\bm r$ the position vector with
respect to the center of the domain and $\bm u$ and $\bm B$ the
velocity and magnetic field vector, respectively. Through integration
by parts these quantities can also be expressed as a function of the
streamfunction $\psi=\nabla^{-2}\omega$ and vector potential
$a=\nabla^{-2}j$, respectively, with $\bm j=j\bm e_z=\nabla\times \bm B$ the current density and $\bm \omega=\omega\bm e_z=\nabla\times \bm u$, the vorticity:
\begin{equation}\label{eqLuPsi}
L_u=2\int_{\Omega} \psi~dA,~~~L_B=2\int_{\Omega}a~ dA,
\end{equation}
in which $a$ and $\psi$ are chosen to be zero at the wall.

A large value of the angular momentum can generally be associated with the presence of a large-scale vortical structure. By analogy we can anticipate that a large value of $L_B$ corresponds to a large-scale current density structure, and we baptize the quantity $L_B$ \emph{angular field}. The evolution equations for $L_u$ and $L_B$ can be derived following the procedure described in Maassen  \cite{Maassen2000}, by time deriving equations (\ref{eqLu}) and using the MHD equations:
\begin{eqnarray} 
\label{qte_mvt}
\frac{\partial{\bm{u}}}{\partial{t}}+(\bm{u}\cdot\nabla)\bm{u}= -\nabla p + \bm{j}\times\bm{B} + \nu\nabla^2\bm{u}\\
\label{eq_magnetic}
 \frac{\partial{\bm{B}}}{\partial{t}} = \nabla\times(\bm{u}\times\bm{B}) + \eta\nabla^2\bm{B}
\end{eqnarray}
together with $\nabla\cdot \bm{u} = 0$ and $\nabla\cdot\bm{B}=0$. The pressure is denoted by $p$, and $\nu$ and $\eta$ are the kinematic viscosity and magnetic diffusivity, respectively. If we write the Lorentz force in the form:
\begin{equation}
\bm j \times \bm B=-\frac{1}{2}\nabla B^2+(\bm B\cdot \nabla)\bm B,
\end{equation}
we can absorb the first term into the pressure term of the Navier-Stokes equations by introducing the modified pressure $p^*=p+B^2/2$. The $(\bm B\cdot \nabla)\bm B$ term does not induce new terms in the equation for $L_u$. It vanishes in a similar way as the
nonlinear term $(\bm u\cdot \nabla)\bm u$ does, using $\nabla\cdot \bm B=0$ and
$\bm B \cdot \bm n|_{\partial\Omega}=0$. The equation for $L_u$ becomes 
\begin{eqnarray}\label{eqdLu}
\frac{d L_u}{d t} =\nu\oint_{\partial {\Omega}}\omega(\bm r\cdot {\bm
  n})ds+ \oint_{\partial {\Omega}}p^*\bm r
\cdot d\bm s.
\end{eqnarray}
The only difference with respect to the hydrodynamic case  \cite{Clercx2001} is  the pressure which is now replaced by the modified pressure $p^*$.
In most fusion plasmas, the quantity $\beta=p/B^2\ll 1$ to insure confinement,
which means that the magnetic part of the pressure dominates. It is important to note that the pressure term in equation (\ref{eqdLu}) vanishes in axi-symmetric domains. In this work we therefore consider both a
circular and a square domain to analyze the influence of this term.

The derivation of the equation for $L_B$ is analogous to the derivation for $L_u$. The resulting equation is:
\begin{eqnarray}
\frac{dL_B}{dt}=\eta\oint_{\partial {\Omega}}j(\bm r\cdot {\bm
  n})ds-2\eta I.\label{dLb}
\end{eqnarray}
We observe that there is a term involving the net current $I$ through
the domain defined by $I=\int_{\Omega} j\bm e_z ~dA$. This term is the
equivalent of the circulation in the hydrodynamic case, which is zero
due to the no-slip walls. The net current is however not imperatively
zero as the tangential magnetic field does not vanish at the
wall. Nevertheless, 
a net current will not be generated if it is initially zero, which is the case in the present work.

We performed computations in two different geometries: a square of
size $D=2$ and a circular geometry with a diameter $D=2.24$. A
description of the generation of the initial conditions and the
numerical scheme, a spectral method with volume penalization, are
given in [\onlinecite{Neffaa2008}]. The initial velocity and magnetic
field consist of correlated Gaussian noise with vanishing
cross-helicity $\int_{\Omega}\bm u\cdot \bm B ~dA$. The magnetic Prandtl number, $\nu/\eta$ is equal to one. The initial Reynolds number, based on the domain size is $\sqrt{2 E_u} D/\nu$ and yields $1960$. The ratio of the magnetic and kinetic energy $E_B/E_u=2.3$, with $E_u = \frac{1}{2}\int_{{\Omega}}\vert\textbf{u}\vert^2~dA$ and $E_B = \frac{1}{2}\int_{{\Omega}}\vert\textbf{B}\vert^2~dA$.
The resolution of the simulations is $512^2$ Fourier modes. In each
geometry 10 runs were performed starting from different statistical
realizations with the same initial parameters. The numerical value of $a$ and $\psi$ is not automatically zero at the domain boundary. This is accomplished \emph{a posteriori} by substracting a constant value at each point in the domain.

In figure \ref{visu} snapshots of the streamfunction and vector
potential  are shown at $t^*=0.75,~3,~12$ with $t^*=t\sqrt{2 E_u(t=0)}/D$. It can be inferred from (\ref{eqLuPsi}) that these quantities should give a good visual interpretation of the presence of angular momentum and field. At time-instant $t^*=0.75$, in which inertial effects are dominant over viscous effects, it is well visible that the velocity field self-organizes into a large domain-filling structure in the square geometry, whereas in the circular geometry several structures are observed. At $t^*=3$ a large structure appears also in the magnetic field in the square geometry. At $t^*=12$ the large-scale velocity and magnetic structures in the square domain are (anti-)aligned. In the circular domain the tendency to create domain filling structures is weaker, even though the magnetic field in the circular domain shows some evidence of the formation of a large current-structure at $t^{*}=12$. To characterize the relaxation of the flows in both geometries, we also show in figure \ref{visu} the decay of the kinetic and magnetic energy in both domains, as well as the absolute value of the cosine of the alignment angle. A continuous decrease of kinetic and magnetic energy is observed and a continuous increase of global alignment.

\begin{figure}
\setlength{\unitlength}{1.\textwidth}
\includegraphics[width=0.4\unitlength]{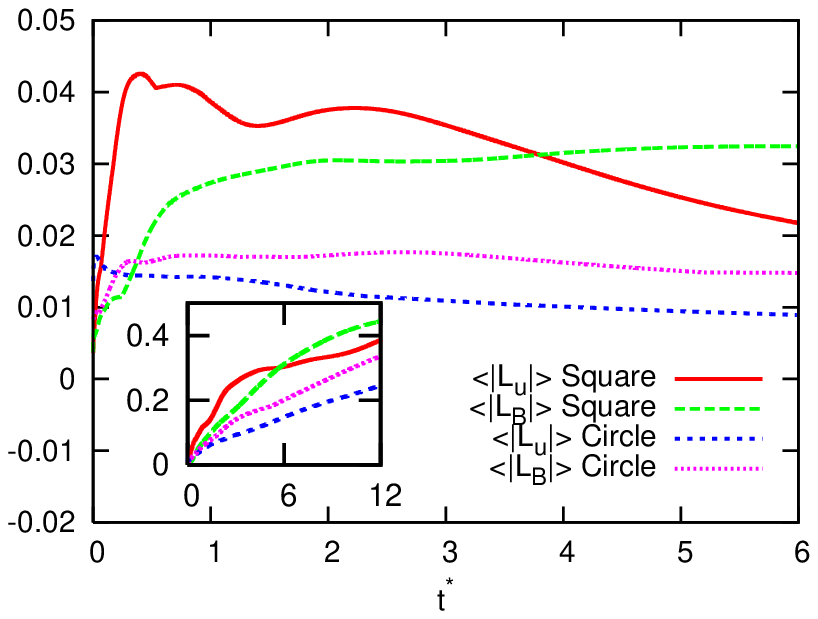}\\
\hspace{-0.02\unitlength}\includegraphics[width=0.46\unitlength]{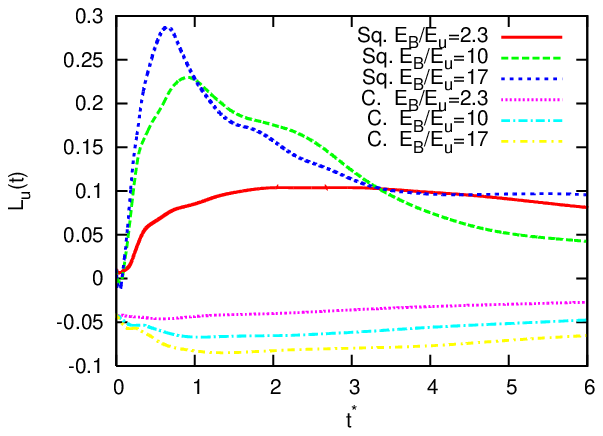}
\caption{(Color online) Top: Time evolution of the absolute value of the angular momentum and angular field, averaged over all realizations, normalized by $\mathcal{L}_u(0)$ and $\mathcal{L}_B(0)$, respectively. In the insert the same quantities are given, normalized by $\mathcal{L}_u(t)$ and $\mathcal{L}_B(t)$ (defined in the text). Bottom: time-dependence of the angular momentum $L_u$ in the square and circular geometry, normalized by $\mathcal{L}_u(0)$. The influence of the magnetic pressure on the spin-up in the square container is illustrated by changing the ratio $E_B/E_u$, while keeping $E_u$ fixed.
\label{figLu}}
\end{figure}

At this moderate Reynolds number, spin-up, \emph{i.e.}, spontaneous
generation of angular momentum, does not occur in every flow
realization. Also the criterion what is strong or weak spin-up is
rather arbitrary. We therefore focus first on mean quantities to
illustrate the general tendency to spin-up. In figure \ref{figLu} we
show the absolute value of the angular momentum, averaged over 10
runs. We take the absolute value because there is no preferential
direction of the spin-up so that an average of the angular momentum
would yield values close to zero for all cases. The time-evolution of
$\langle|L_u|\rangle$ and $\langle|L_B|\rangle$ is shown for both the square and the circular geometry. $\langle\cdot\rangle$ denotes the average over $10$ realizations. The quantities are normalized by 
$\mathcal{L}_u(0)=\|r\|_2\sqrt{2\langle E_u(t=0)\rangle}$ and $\mathcal{L}_B(0)=\|r\|_2\sqrt{\langle 2E_B(t=0)\rangle}$, with $\|r\|_2$ the Euclidean norm of $\bm r$. The quantity $\mathcal{L}_u(t)$ corresponds to the value of the angular momentum of a flow in solid-body rotation with kinetic energy $\langle E_u(t)\rangle$, which is the flow which optimizes the value of the angular momentum for a given kinetic energy.  By analogy, $\mathcal{L}_B(t)$ is used to normalize the angular field. The following is observed: at short times $L_u$ rapidly increases in the square, but does not increase in the circular geometry. The value of $L_B$ also increases in the square, but delayed with respect to $L_u$. In the circular geometry an increase of $L_B$ is also observed. In the inset the values of $\langle |L_u|\rangle$ and $\langle |L_B|\rangle$ are plotted normalized by $\mathcal{L}_u(t)$ and $\mathcal{L}_B(t)$. 
This normalization has the advantage to correct for the decay of the kinetic and magnetic energy but has the disadvantage that it is sensitive to selective decay \cite{KeetelsPhD} so that at long times we observe generation of angular momentum in each case even if its absolute value might be small. 
In the following we will give, where possible, an explanation for the $4$ curves in figure \ref{figLu}.

First, in the square geometry a strong spin-up of the velocity field is observed. In the hydrodynamic case, it was argued in [\onlinecite{Clercx2001,Keetels2008}] that the pressure term triggers the spin-up in the square geometry. The magnetic field enhances the pressure term through the magnetic pressure ($p^*=p+B^2/2$). If in the present case it is also the pressure term in (\ref{eqdLu}) which triggers the spin-up, the effect could be enhanced by increasing the magnetic fluctuation strength $B^2$. This is illustrated in figure \ref{figLu} (bottom). For one run in which spin-up was observed,  the initial magnetic fluctuations  are increased from $E_B/E_u=2.3$ up to $E_B/E_u=10$ and $16.7$, while keeping the initial $E_u$ fixed. The resulting spin-up is significantly stronger.

Second, for $L_u$ in the circular geometry, like in the hydrodynamic case \cite{Schneider2005-2}, no spontaneous spin-up is observed. Increasing the magnetic-field strength does only weakly influence this result (figure \ref{figLu}, bottom).

Third, the interpretation of the generation of the angular field in the square geometry is less straightforward, as equation (\ref{dLb}) does not contain a pressure term.  
The tendency to create large-scale magnetic structures can be
attributed to the selective decay mecanism \cite{Matthaeus1980}, which
was recently shown to persist in bounded geometries
\cite{Neffaa2008}. This does however not explain the symmetry breaking
or angular momentum generation, which is the main issue of the present
work. A possible trigger for the spin-up could be alignment.
It is well known that the magnetic field and the velocity field tend to align 
so that the non-linear term in the equation for $j$ (or $\bm B$)
vanishes. Hence, the magnetic field tends to an alignment with the
velocity field which acquired angular momentum through the modified
pressure term. It is therefore expected that the magnetic spin-up
follows the hydrodynamic spin-up after a time-scale corresponding to
the alignment. Indeed, $L_B$ spins-up shortly after $L_u$. The cosine of the angle between $\bm u$ and $\bm B$, measuring the global alignment, is plotted in the inset of figure \ref{visu} (right). A tendency towards global alignment is observed for long times.

Fourth, in the circular geometry, the weak spin-up of the magnetic field is surprising. Higher resolution simulations are needed to clarify whether this is a viscous effect and/or a statistically more probable (maximum entropy) state? In this context we can refer to \cite{Taylor2008}, where, based on point-vortices, it was shown that two types of most probable states exist in a circular domain: a double vortex, free from angular momentum and an axi-symmetric flow, with finite angular momentum. This work neglected the influence of viscosity so that it is not clear how the angular momentum is acquired in the circular geometry.

We now resume our findings. Rapid generation of angular momentum takes place in bounded MHD turbulence, as long as the geometry is non-axisymmetric. The effect is enhanced by the magnetic pressure. On a slower time-scale also magnetic spin-up is observed in both geometries. It is not clear how this angular field is created. Both alignment and selective decay could be possible explanations. 

We want to stress the implications of the present study for
confinement research. Fusion plasmas are wall bounded and not
axi-symmetric, so that even in the case of charge neutrality the
plasma might have a tendency to create zonal flows and zonal fields, depending on the geometry of the cross-section of the plasma and the strength of the magnetic fluctuations.  The present work opens several perspectives for future research, such as the influence of $Pr_m$, $Re$, and in particular the extension to three dimensions in which the effects of imposed magnetic fields, currents and toroidal velocities can be taken into account.

We acknowledge valuable discussions with Herman Clercx, David Montgomery and Geert Keetels. This work was supported by the \emph{ANR} under the
 contract \emph{M2TFP}.

\end{document}